\documentclass[12pt]{article}
\usepackage{graphicx}
\usepackage{subcaption}
\usepackage{amsmath,amssymb,xcolor,physics}
\usepackage{url}
\usepackage{ulem}
\usepackage{ascmac}
\usepackage{caption}
\usepackage{comment}
\usepackage{xcolor}
\usepackage{hyperref}

\usepackage{tabularray}
\usepackage[export]{adjustbox}

\allowdisplaybreaks[4]

\begin{document}


\begin{center}
{\Large Exponential improvement in quantum simulations of bosons}
\end{center}
\vspace{0.1cm}
\vspace{0.1cm}
\begin{center}

Masanori Hanada,$^{a}$ Shunji Matsuura,$^{b}$ Emanuele Mendicelli,$^{c}$ and Enrico Rinaldi$^{d}$

\end{center}
\vspace{0.3cm}
{\small
\begin{center}
$^{a}$
School of Mathematical Sciences, Queen Mary University of London\\
Mile End Road, London, E1 4NS, United Kingdom\\
\vspace{1mm}
$^a$qBraid Co., Harper Court 5235, Chicago, IL 60615, United States\\
\vspace{1mm}
$^{b,d}$RIKEN Center for Interdisciplinary Theoretical and Mathematical Sciences(iTHEMS),\\
RIKEN, Wako, Saitama 351-0198, Japan\\
\vspace{1mm}
$^b$Department of Electrical and Computer Engineering,
University of British Columbia\\
Vancouver, BC V6T 1Z4, Canada\\
\vspace{1mm}
$^b$Department of Physics, University of Guelph, ON N1G 1Y2, Canada \\
\vspace{1mm}
$^b$Center for Mathematical Science and Advanced Technology\\
Japan Agency for Marine-Earth Science and Technology,
Yokohama 236-0001, Japan \\
\vspace{1mm}
$^{c}$
Department of Mathematical Sciences, University of Liverpool\\ 
Liverpool L69 7ZL, United Kingdom\\
\vspace{1mm}
$^d$Quantinuum K.K., Otemachi Financial City Grand Cube 3F\\
1-9-2 Otemachi, Chiyoda-ku, Tokyo, Japan
\\
\vspace{1mm}
$^{d}$Center for Quantum Computing (RQC), RIKEN,
Wako, Saitama 351-0198, Japan\\
\vspace{1mm}
$^{d}$Theoretical Quantum Physics Laboratory, Cluster of Pioneering Research, RIKEN\\
Wako, Saitama 351-0198, Japan\\
\end{center}
}


\newpage 

\begin{center}
  {\bf Abstract}
\end{center}
Hamiltonian quantum simulation of bosons on digital quantum computers requires truncating the Hilbert space to finite dimensions. 
The method of truncation and the choice of basis states can significantly impact the complexity of the quantum circuit required to simulate the system.
For example, a truncation in the Fock basis where each boson is encoded with a register of $Q$ qubits, can result in an exponentially large number of Pauli strings required to decompose the truncated Hamiltonian. 
This, in turn, can lead to an exponential increase in $Q$ in the complexity of the quantum circuit. 
For lattice quantum field theories such as Yang-Mills theory and QCD, several Hamiltonian formulations and corresponding truncations have been put forward in recent years. 
There is no exponential increase in $Q$ when resorting to the orbifold lattice Hamiltonian, while we do not know how to remove the exponential complexity in $Q$ in the commonly used Kogut-Susskind Hamiltonian.
Specifically, when using the orbifold lattice Hamiltonian, the continuum limit, or, in other words, the removal of the ultraviolet energy cutoff, is obtained with circuits whose resources scale like $Q$, while they scale like $\mathcal{O}(\exp(Q))$ for the Kogut-Susskind Hamiltonian: this can be seen as an exponential speed up in approaching the physical continuum limit for the orbifold lattice Hamiltonian formulation.
We show that the universal framework, advocated by three of the authors (M.~H., S.~M., and E.~R.) and collaborators, provides a natural avenue to solve the exponential scaling of circuit complexity with $Q$, and it is the reason why using the orbifold lattice Hamiltonian is advantageous.
We also point out that Hamiltonian formulations based on a gauge-invariant Hilbert space require an exponential increase in the resource requirement with respect to using an extended Hilbert space. Specifically, for $3+1$ dimensional theories, the qubit count could be reduced by about 30\% while the depth and compilation cost increase exponentially, diminishing the hope for any sort of quantum advantage.
 
\newpage
\tableofcontents
\section{Introduction}
Certain tasks and problem formulations are better suited for classical computers, while others are more appropriate for quantum computers. 
What types of tasks or problem formulations are best suited for quantum computers? 
These would include problems that benefit from simple quantum algorithms, where significant speedup and scalability can be achieved, and where classical pre- and post-processing do not become a bottleneck.
While both classical and quantum computers are valuable tools, depending on tasks or formulations, we should use the better one.

When we use computers, either classical or quantum, the choice of basis matters because we may or may not be able to design simple algorithms (or quantum circuits), and despite physics not caring about the basis, we physicists do.
A significant fraction of our job is to find a good basis; for example, if we can find a simple way to write the energy eigenbasis for a quantum system of interest, we do not even need a quantum computer!

For quantum simulation of quantum many-body physics, while the number of qubits required is almost always discussed, the gate counting is sometimes ignored, as it can be highly dependent on the simulation algorithm as well as the basis in which one chooses to work.
Producing resource estimates for quantum algorithms in quantum chemistry or quantum simulations of condensed matter systems and lattice gauge theories is a challenging task that has been studied extensively in the literature~\cite{Lee:2020egw,Yoshioka:2022rej}.
As long as `typical' spin or fermion systems are discussed, like the ones describing quantum magnetism or quantum chemistry, the Hamiltonian can often be directly encoded into qubit operators and often expressed in terms of Pauli operator strings: the number of gates does not increase exponentially for standard tasks such as Hamiltonian time evolution.
The reason is that there is often a natural or intuitive mapping between the system under study and the digital quantum computer: the Hamiltonians are defined in a good basis that requires only a polynomial number of Pauli terms.

The situation is different for the simulation of bosons on digital quantum computers with qubits. 
Because the Hilbert space of the systems with bosons is infinite-dimensional, we need to truncate it and write the Hamiltonian in terms of quantum gates acting on qubits. 
Unless we choose an appropriate truncation scheme, the truncated Hamiltonian can be complicated and gate counting can increase dramatically, leading to an asymptotic exponential number of gates with respect to the number of qubits $Q$ used for truncating the local Hilbert space of each bosonic degree of freedom.
\footnote{Note that, even for spin systems, we could perform a complicated unitary transformation and write the Hamiltonian in a complicated form. If we do not know the unitary transformation that takes the Hamiltonian back to the original simple form, we cannot design an efficient circuit.}
This problem appears in a few standard treatments of bosonic systems which are currently employed in the Hamiltonian formulations of lattice gauge theories, for example using the Kogut-Susskind Hamiltonian~\cite{Kogut:1974ag,Byrnes:2005qx}.
When looking at the long term goal of simulating quantum field theories on a fault-tolerant quantum computer, this exponential growth in the number of gates should be addressed, because the number of qubits $Q$ needed to encode a bosonic degree of freedom is expected to increase as we approach the lattice continuum limit, or in other words, as we remove the ultraviolet cutoff.

In this work, we will show that we can easily avoid the exponential in $Q$ gate requirement using quantum Fourier transform~\cite{Coppersmith:2002skh}, which connects the coordinate and momentum bases. 
This method allows us to write simple quantum circuits using the universal framework for the simulation of bosonic systems introduced in Ref.~\cite{Halimeh:2024bth} by three of the authors of this study (M.~H., S.~M., and E.~R.).
Note that the universal framework includes fundamental quantum field theories such as Yang-Mills theory and Quantum Chromodynamics (QCD), both based on the orbifold lattice Hamiltonian~\cite{Buser:2020cvn,Bergner:2024qjl,Kaplan:2002wv}. 
Therefore, we can simulate Yang-Mills theory and QCD without suffering from the exponential growth in the number of quantum gates by adopting the orbifold lattice.
We point out that we currently do not see a straightforward way to circumvent this problem for the Kogut-Susskind Hamiltonian, due to the lack of a quantum Fourier transform.
It is possible that future research and the use of sophisticated quantum oracles that require detailed group-theoretic considerations could lead to efficient circuit design for the Kogut-Susskind Hamiltonian~\cite{Kan:2021xfc,Rhodes:2024zbr}.
However, designing quantum circuits in terms of native gates is a challenging task that may demand exponentially large classical resources~\cite{Balaji:2025afl}, as discussed in Sec.~\ref{sec:when_does_it_matter}. 
Although we believe that a compiler would handle the generation of quantum circuits for any given algorithm, this task is in general exponentially hard, and compiling can become the bottleneck for quantum simulation. 
Naturally, we cannot exclude the possibility of an efficient way to encode the group-theoretic manipulations necessary for the Kogut-Susskind Hamiltonian, but finding such a new method would require a significant breakthrough in analytic understanding, which at least has to remove the necessity of exponentially large input for classical computations.
Some headway in this direction might come from studying different limits of the Kogut-Susskind Hamiltonian~\cite{Ciavarella:2024fzw,Ciavarella:2025bsg}.

This paper is organized as follows. Section~\ref{sec:theoretical_setup} introduces a class of Hamiltonians, schematically written as Eq.~\eqref{generic_Hamiltonian}, that enables efficient quantum simulations. For these Hamiltonians, efficient quantum circuits can be designed explicitly, as demonstrated in numerous studies [cite relevant papers including Halimeh et al]. Section~\ref{sec:universal_framework} reviews why this class of Hamiltonians can be efficiently simulated on fault-tolerant digital quantum computers. Although these simulation protocols are well established, their true advantages may not be fully appreciated. Although the qubit requirements for implementation are routinely discussed, gate complexity receives less attention. We argue that the primary advantage of these protocols lies in their circuit simplicity, which encompasses both low gate counts and negligible compilation costs. Because quantum circuit design can be handled analytically for most tasks, the computational resources required for circuit compilation are minimal.
Section~\ref{sec:inefficiency} illustrates this point by examining alternative, inefficient methods. While one might expect to simply avoid inefficient approaches, this is not always feasible. Section~\ref{sec:when_does_it_matter} presents several cases where efficient protocols are unavailable, including the Kogut-Susskind Hamiltonian for non-Abelian gauge theory. Since the orbifold lattice Hamiltonian belongs to the class defined in eq.~\eqref{generic_Hamiltonian}, we reach a significant conclusion: the orbifold lattice Hamiltonian provides exponential speedup for quantum simulations of non-Abelian gauge theory compared to the best known estimates for approaches based on the Kogut-Susskind Hamiltonian. Sec.~\ref{sec:discussion} discusses the consequences of our findings.

\section{Theoretical setup}\label{sec:theoretical_setup}
In this paper, we will mostly be concerned with Hamiltonians of the form
\begin{align}
\hat{H}
=
\sum_{a=1}^B\frac{\hat{p}_a^2}{2}
+
V(\hat{x}_1,\cdots,\hat{x}_B)\, 
\label{generic_Hamiltonian}
\end{align}
where $\hat{x}_a$ and $\hat{p}_a$ ($a=1,\cdots,B$) are coordinate and momentum operators, respectively, which satisfy the canonical commutation relation 
\begin{align}
[\hat{x}_a,\hat{p}_b]=\mathrm{i}\delta_{ab}\, . 
\end{align}
We assume $V(\hat{x}_1,\cdots,\hat{x}_B)$ is a polynomial of degree $d\ge 4$, because we are interested in interacting theories whose energies are bounded from below.
Here we list a few examples:
\begin{itemize}
\item If we have $n_{\rm particle}$ bosonic particles in three-dimensional space, then $B=3n_{\rm particle}$,
\item For a scalar QFT on a lattice with $n_{\rm lattice}$ points (see, e.g., Ref.~\cite{Jordan:2012xnu}), there is a bosonic degree of freedom at each lattice site and therefore $B=n_{\rm lattice}$, 
\item For the orbifold lattice Hamiltonian for SU($N$) Yang-Mills theory with $n_{\rm link}$ links, there are $2N^2$ bosonic degrees of freedom on each link, and hence $B=2N^2n_{\rm link}$.
\end{itemize}

To truncate the Hilbert space, we assign $Q$ qubits for each boson. 
The total number of qubits is $BQ$ and the dimension of the Hilbert space is $2^{BQ}$. 
The Hilbert space corresponding to each boson has the dimension $\Lambda=2^Q$. 
To recover the original theory without truncation, we take the limit of $Q\to\infty$. 
In this paper, we focus on the scaling of the computational cost with respect to $Q$, specifically the complexity of the quantum circuits.

As we will see in Sec.~\ref{sec:universal_framework}, using the universal framework of Ref.~\cite{Halimeh:2024bth}, the number of quantum gates in the circuits scales as a polynomial of $Q$.
On the other hand, when the quantum Fourier transform is not employed, we observe the exponential growth of the number of gates, as we will see in Sec.~\ref{sec:inefficiency}.
The examples we consider are governed by Hamiltonians that look very simple, as in ~\eqref{generic_Hamiltonian}.
Nevertheless, they are able to demonstrate the power of the quantum Fourier transform, one of the best-known subroutines in quantum computing, in exponentially reducing the complexity of quantum circuits with respect to other algorithms.

The scaling of the gate complexity discussed in Sec.~\ref{sec:universal_framework} and 
Sec.~\ref{sec:inefficiency} is not surprising. 
The nontrivial question is: \textit{when does this scaling matter?} 
As we will see in Sec.~\ref{sec:when_does_it_matter}, there are important theories including Yang-Mills theory and QCD in which the value of $Q$ needed for practical applications can increase with the number of bosons, and different complexities -- exponential vs polynomial -- can make a difference. 
For the study of Yang-Mills theory and QCD, this universal framework based on the quantum Fourier transform is applicable if we use the orbifold lattice Hamiltonian which has the simple form~\eqref{generic_Hamiltonian}. 
However, it does not apply to the Kogut-Susskind Hamiltonian due to the lack of an efficient algorithm for the quantum Fourier transform.\footnote{
See Ref.~\cite{Murairi:2024xpc} for the study of the quantum Fourier transform on a special kind of discrete subgroup of SU(3).}
We see this as a further motivation to use the orbifold lattice construction for lattice gauge theory simulations in the Hamiltonian formalism. 
\section{Polynomial efficiency of universal framework}\label{sec:universal_framework}
In this section, we review the universal framework~\cite{Halimeh:2024bth} and recall why the number of Pauli strings scales polynomially with the number of qubits, $Q$. 
This framework uses the coordinate basis and momentum basis connected via the quantum Fourier transform. 
The word `universal' is used in this context because, despite the simplicity, this framework applies to a wide class of theories of interest, including Yang-Mills theory and QCD, which is highly nontrivial.
\subsection{Coordinate basis}\label{sec:coordinate_basis}
First, we define the coordinate basis that is convenient for dealing with the potential term of the Hamiltonian of \eqref{generic_Hamiltonian}. 
For past works on the coordinate basis, see, e.g., refs.~\cite{Macridin:2018oli, Macridin:2018gdw, Klco:2018zqz, Macridin:2021uwn, Chandra:2022mae,Li:2022ped, Hanada:2022pps}.

We have $B$ bosonic coordinate variables $\vec{x}=(x_1,\cdots,x_{B})$. 
For each of them, we can define the Hilbert space $\mathcal{H}_a$ spanned by the coordinate eigenstate $\ket{x_a}$ that satisfies $\hat{x}_a\ket{x_a}=x_a\ket{x_a}$:
\begin{align}
\mathcal{H}_a
=
\mathrm{Span}\{\ket{x_a}|x_a\in\mathbb{R}\}\, . 
\end{align}
For the system of all bosons, we have the coordinate eigenstate
\begin{align}
\ket{\vec{x}}
=
\otimes_a\ket{x_a}\, ,
\end{align}
and the full Hilbert space
\begin{align}
\mathcal{H}
=
\otimes_a\mathcal{H}_a\, . 
\end{align}
To truncate the Hilbert space to finite dimensions, we compactify each $x_a$ to a large circle of circumference $2R$, with periodic boundary condition $x\sim x+2R$, and approximate the circle by $\Lambda$ points uniformly distributed on the circle.\footnote{
We chose periodic boundary conditions to make the Fourier transform straightforward. Note that despite this discretization choice being reminiscent of using discrete subgroups, it leads to very different algorithms. 
} 
The spacing between the points is
\begin{align}
   \delta_x=\frac{2R}{\Lambda}\, ,  
\end{align}
and the points are taken to be
\begin{align}
    x_{a,n_a}
    =
    -\frac{\Lambda-1}{\Lambda}R + n_a\delta_x
    =
    \left(n_a-\frac{\Lambda-1}{2}\right)\delta_x\, , 
    \qquad
    n_a=0,1,\cdots,\Lambda-1\, . 
\end{align}
Hence, $x_{a,n_a}$ takes values $\pm\frac{\delta_x}{2}$, $\pm\frac{3\delta_x}{2}$, ..., $\pm\frac{(\Lambda-1)\delta_x}{2}$. 
This specific choice manifests the symmetry under $x \leftrightarrow -x$ that exists for many theories, and it can be seen as a discretized grid of points.

To reduce the truncation effect, we must take $R$ large enough for $x\sim\pm R$ not to be significantly excited, such that the boundary condition does not affect the physics of interest.
In other words, the corresponding eigenvector $\ket{x_a}$ does not have a large amplitude.
We assign $Q$ qubits for each boson and take $\Lambda=2^Q$. 
Using a binary mapping with $b_{a,i}=\{0,1\}$, 
\begin{align}\label{eq:binary_mapping}
\ket{x_{a,n_a}} =\ket{b_{a,1}}\ket{b_{a,2}}\cdots\ket{b_{a,Q}}\, , 
\qquad
n_a=b_{a,1}+2b_{a,2}\cdots+2^{Q-1}b_{a,Q}\, ,
\end{align}
and the Pauli Z operator defined by 
\begin{align}
    \hat{\sigma}_z \equiv \ket{0}\bra{0} - \ket{1}\bra{1}\, , 
    \label{eq: pauli Z}
\end{align}
we can write the coordinate operator $\hat{x}_a$ as 
\begin{align}
\hat{x}_a
=
-
\delta_x\cdot
\left(
\frac{\hat{\sigma}_{z;a,1}}{2}
+
2\cdot\frac{\hat{\sigma}_{z;a,2}}{2}
+
\cdots
+
2^{Q-1}\cdot\frac{\hat{\sigma}_{z;a,Q}}{2}
\right)\, . 
\end{align}
Here $\hat{\sigma}_{z;a,j}$ is the Pauli Z operator acting on the $j$-th qubit assigned to the $a$-th boson. 

In the coordinate basis, each interaction term between $d$ bosons can be written as a sum of $Q^d$ Pauli strings made of $d$ Pauli $\hat{\sigma}_z$ gates.
\subsection{Quantum Fourier transform and momentum basis}\label{sec:momentum_basis}
By applying the Fourier transform, we can switch to momentum eigenstates. 
To keep the symmetry under $p\leftrightarrow -p$, $x\leftrightarrow -x$ that exists in many theories, we can take 
\begin{align}
    p_{a,\tilde{n}_a}
    =
    \left(\tilde{n}_a-\frac{\Lambda-1}{2}\right)\delta_p\, , 
    \qquad
    \delta_p=\frac{\pi}{R}\, , 
    \qquad
    \tilde{n}_a=0,1,\cdots,\Lambda-1\, . 
\end{align} 
We can choose the Fourier Transform as 
\begin{align}
    \ket{p_{a,\tilde{n}_a}}
    =
    \frac{1}{\sqrt{\Lambda}}\sum_{x_{a,n_a}} e^{\mathrm{i}p_{a,\tilde{n}_a}x_{a,n_a}}\ket{x_{a,n_a}}\, . 
\end{align}
Fourier transform can be realized efficiently on quantum computers as \textit{quantum Fourier transform}~\cite{Coppersmith:2002skh}, with an order $Q^2$ number of quantum gates for the textbook algorithm.
Improved algorithms also exist reducing the cost to order $Q \log{Q}$ when an approximation can be tolerated~\cite{Hales:2000hls}.

We can define $\hat{p}_a$ as
\begin{align}
    \hat{p}_a\ket{p_{a,\tilde{n}_a}}
    =
    p_{a,\tilde{n}_a}
    \ket{p_{a,\tilde{n}_a}}\,  , 
    \label{p-option-2}
\end{align}
In this basis, we can write the momentum operator $\hat{p}_a$ as 
\begin{align}
\hat{p}_a
=
-
\delta_p\cdot
\left(
\frac{\hat{\sigma}_{z;a,1}}{2}
+
2\cdot\frac{\hat{\sigma}_{z;a,2}}{2}
+
\cdots
+
2^{Q-1}\cdot\frac{\hat{\sigma}_{z;a,Q}}{2}
\right)\, . 
\end{align}
Each $\hat{p}_a^2$ contains $\frac{Q(Q-1)}{2}$ nontrivial Pauli strings of length two. 
In this paper, because we are interested in the asymptotic scaling at large $Q$, we will not distinguish $Q(Q-1)$ and $Q^2$. 
\subsection{Gate counting of various algorithms}\label{sec:gate_counting}
Our universal framework uses the coordinate basis for the potential term and the momentum basis for the kinetic part, connecting two bases via the quantum Fourier transform.
The gate counting is straightforward, as demonstrated in Ref.~\cite{Halimeh:2024bth} for the Hamiltonian time evolution via Trotter-Suzuki decomposition~\cite{Trotter:1959, Suzuki:1976be}.
While the states in those bases are represented using a binary mapping for simplicity (see Eq.~\eqref{eq:binary_mapping}), it is still an open question how the gate counting changes when using different encodings or when using a set of basis functions instead of a simple grid to discretize the Hilpert space.

As before, we consider a system of $B$ bosonic degrees of freedom and encode each of them using $Q$ qubits, or equivalently, using a $2^Q$-dimensional Hilbert space. 
Therefore, the dimension of the full Hilbert space is $2^{BQ}$. 
We assume the interaction consists of $d$-boson couplings, with $C$ terms in total. 
For simplicity, we also assume $C>B$ and $d\ge 4$, which is a rather standard situation, e.g. in lattice models with nearest-neighbours interactions. 
Based on these assumptions, we show the scaling of the number of gates needed for quantum simulations with this framework, focusing on the asymptotic behavior as $Q\to\infty$.

As we demonstrate below, the number of gates needed for quantum simulations in the universal framework scales polynomially in $Q$. 
On the other hand, as we will see in later sections, many other approaches lead to a quick growth of the number of gates that is exponential in $Q$. 
As we will discuss in Sec.~\ref{sec:when_does_it_matter}, such exponential growth can make quantum simulation infeasible for a certain class of theories.

\subsubsection{Hamiltonian time evolution via Lie-Trotter decomposition}\label{sec:time_evolution}

As a first look into the problem of the scaling of resources with $Q$, we consider the Hamiltonian time evolution within the standard Lie-Trotter decomposition.
In this short section we follow the results of Ref.~\cite{Halimeh:2024bth}, and summarize the salient points which we will need for the later discussion of the exponential inefficiency of other approaches.

For simplicity, we use the simplest setup without higher-order improvement, with a Trotter time step $\Delta t$. 
We can perform infinitesimal time evolution $\Delta t \to 0$ alternating between the potential and the kinetic parts, inserting the quantum Fourier transform in between them at every step. 

One Trotter step can be realized as follows: 
\begin{itemize}
    \item
    Act with $\exp\left(-\mathrm{i}V(\hat{x}_1,\cdots,\hat{x}_B)\Delta t\right)$ on the quantum state expressed in the coordinate basis. 
    This operator can be written as a product of terms corresponding to $CQ^d$ Pauli strings as shown in Sec.~\ref{sec:coordinate_basis}. 
    Because the lengths of Pauli strings are at most $d$ and do not increase with $Q$, the number of gates needed for this part is proportional to $CQ^d$.

    \item 
    Perform a quantum Fourier transform to switch to the momentum basis, which requires $\sim BQ^2$ gates~\cite{Coppersmith:2002skh}. 
    
    \item
    Act with $\exp\left(-\frac{\mathrm{i}}{2}\sum_a\hat{p}_a^2\Delta t\right)$ on the quantum state expressed in the momentum basis. 
    This operator can be written as a product of the terms corresponding to the $BQ^2$ Pauli strings as shown in Sec.~\ref{sec:momentum_basis}. 
    Because the lengths of Pauli strings do not scale with $Q$ (specifically, it is at most $2$), the number of gates needed for this part is proportional to $BQ^2$.

    \item 
    Perform an inverse quantum Fourier transform to switch back to the coordinate basis for the next step of the time evolution.
    This again requires $\sim BQ^2$ gates. 

\end{itemize}
The dominant part is $\exp\left(-\mathrm{i}V(\hat{x}_1,\cdots,\hat{x}_B)\Delta t\right)$ that requires $\sim CQ^d$ gates.

\subsubsection{Hamiltonian time evolution via quantum signal processing}\label{sec:signal_processing}

While the Trotter decomposition of Hamiltonian simulation on quantum computers is well known and simple to implement, one may wonder if a similar polynomial scaling of resources is achieved when using other more complicated algorithms for achieving the same task.
In this short section we provide another example, where Hamitlonian time evolution is implemented using quantum signal processing~\cite{Low:2016sck} or QSP.

QSP requires the block encoding of Pauli strings, whose cost increases with the number and lengths of Pauli strings.
Although block encoding is a well-known procedure, let us explain the concrete steps explicitly to see how the quantum Fourier transform can be utilized. 
(See Ref.~\cite{Hariprakash:2023tla} that explained this by using the scalar $\phi^4$ QFT as an example.)

In the coordinate basis, our Hamiltonian is written in terms of Pauli strings $\hat{\Pi}_i$, $\hat{\tilde{\Pi}}_{j}$ and the quantum Fourier transforms $\hat{F}$ as
\begin{align}
\hat{H}
=
\sum_{i} \alpha_{i} \hat{\Pi}_{i}
+
\sum_{j} \tilde{\alpha}_{j}
\hat{F}^\dagger \hat{\tilde{\Pi}}_{j}\hat{F}
, \qquad \alpha_{i}>0,\quad \tilde{\alpha}_{j} > 0 
\label{Hamiltonian-as-sum-of-Pauli-chains}
\end{align}
where the first and second terms on the right-hand side are the potential and kinetic terms.

The first step in quantum signal processing is the block-encoding of the Hamiltonian into a unitary matrix. 
We introduce ancilla states $\ket{i}$, $\ket{\tilde{j}}$ and prepare a state $|G\rangle$ defined by  
\begin{align}
\ket{G}  
=  
\sum_i g_i \ket{i}
+
\sum_i \tilde{g}_i \ket{\tilde{j}}\, , 
\end{align}
where
\begin{align}
|g_i|^2
=
\frac{\alpha_i}{\lambda}\, , 
\qquad
|\tilde{g}_j|^2
=
\frac{\tilde{\alpha}_j}{\lambda}\, , 
\qquad
\lambda
=
\sum_i\alpha_i
+
\sum_j\tilde{\alpha}_j\, . 
\end{align} 
Then we can prepare a unitary operator $\hat{U}$ acting on the Hilbert space extended with ancilla qubits, which satisfies 
\begin{align}
\frac{\hat{H}}{\lambda} = \Big( \bra{G} \otimes \hat{I} \Big)  \hat{U}  \Big( \ket{G} \otimes\hat{I} \Big)\, .
\label{eq:H_G_U}
\end{align}
Here, $\hat{I}$ is the identity operator acting on the original Hilbert space.
The unitary operator $\hat{U}$ can be constructed from control-$\hat{\Pi}_i$ and control-$\hat{\tilde{\Pi}}_i$ operations, 
\begin{align}
\hat{U} \ket{i} \ket{\psi}=\ket{i} \left(\hat{\Pi}_i \ket{\psi}\right)\, , 
\qquad
\hat{U} \ket{\tilde{j}} \ket{\psi}=\ket{\tilde{j}} \left(\hat{F}^\dagger\hat{\tilde{\Pi}}_j\hat{F}\ket{\psi}\right)\, . 
\end{align}
More explicitly, 
\begin{align}
\hat{U} 
=
\sum_i
\left(
\ket{i}\bra{i}
\otimes
\hat{\Pi}_i
\right)
+
\hat{F}^\dagger
\sum_j
\left(
\ket{\tilde{j}}\bra{\tilde{j}}
\otimes\hat{\tilde{\Pi}}_j
\right)
\hat{F}\, . 
\end{align}
This can be written as
\begin{align}
\hat{U} 
=
\hat{F}^\dagger
\hat{U}_{\rm kin}
\hat{F}
\hat{U}_{\rm pot}\, ,  
\end{align}
where
\begin{align}
\hat{U}_{\rm pot}
=
\sum_i
\left(
\ket{i}\bra{i}
\otimes
\hat{\Pi}_i
\right)
+
\left(
\sum_j
\ket{\tilde{j}}\bra{\tilde{j}}
\otimes\hat{I}
\right)\, , 
\end{align}
\begin{align}
\hat{U}_{\rm kin}
=
\sum_i
\left(
\ket{i}\bra{i}
\otimes
\hat{I}
\right)
+
\sum_j
\left(
\ket{\tilde{j}}\bra{\tilde{j}}
\otimes\hat{\tilde{\Pi}}_j
\right)\, . 
\end{align}
This unitary is the block encoding of the Hamiltonian. 
The complexity of building it is dominated by the cost for $\hat{U}_{\rm pot}$, which is proportional to $CQ^d$. 
Once we have the block encoding, the rest of quantum signal processing~\cite{Low:2016sck} goes through without change. 
The key point is that, although the block encoding with Pauli strings of the form $\hat{U} 
=
\sum_i
\left(
\ket{i}\bra{i}
\otimes
\hat{\mathcal{O}}_i
\right)$, where $\hat{\mathcal{O}}_i$ are Pauli strings, is usually assumed, the only thing actually used is that $\hat{\mathcal{O}}_i$ is unitary and Hermitian.

The complexity of the simulation is roughly proportional to the cost of block encoding, $CQ^d$.

\section{Exponential inefficiency of other approaches}\label{sec:inefficiency}
This section considers approaches that do not employ the quantum Fourier transform, but are routinely used in the literature, both for resource estimation and for experimental implementations of quantum algorithms.
The goal is to demonstrate that gate counting can increase significantly if we do not use a well-crafted technique. 
As an example, we consider the coordinate basis truncation and the Fock basis truncation, with the Pauli string expansion approach. 
We show explicitly that the number of Pauli strings increases exponentially with $Q$ and quantum circuits become exponentially long.
In fact, in these simple examples, it is possible to improve the circuits by taking different approaches, for example, using qubitization~\cite{Low:2016znh} or randomized compiling~\cite{PhysRevLett.123.070503}.
However, the situation becomes complicated when we consider more nontrivial examples such as the Kogut-Susskind Hamiltonian for SU($N$) gauge theory, and to our knowledge, there is no known way to avoid the exponential increase in the number of gates.
In most realistic cases, like QCD in three spatial dimensions, it is not even possible to count the gates needed to implement the time evolution of the Kogut-Susskind Hamiltonian, unless one already knows all the matrix elements~\cite{Rhodes:2024zbr}.

\subsection{Problem with coordinate basis}\label{sec:naive_coord_basis}
What happens if we do not use the Fourier transform? 
Then exponentially many Pauli strings appear from $\hat{p}^2$.

Because we do not use the Fourier transform, we do not have to use the periodic boundary condition. 
By using notation $\ket{n_a}=\ket{x_{a,n_a}}$, a convenient way of regularizing $\hat{p}_a^2$ is 
\begin{align}
\hat{p}_a^2
=
\frac{1}{\delta_x^2}
\sum_{n_a=0}^{\Lambda-1}
\Bigl(
2\ket{n_a}\bra{n_a}
-
\ket{n_a+1}\bra{n_a}
-
\ket{n_a}\bra{n_a+1}
\Bigl)\, . 
\label{p^2-coordinate-basis}
\end{align}
With the periodic boundary condition, $\ket{\Lambda}=\ket{0}$ is assumed. Without the periodic boundary condition, the terms including $\ket{\Lambda}$ or $\bra{\Lambda}$ should be ignored. 
Below, let us not impose the periodic boundary condition. 
Furthermore, we focus on one boson and drop the subscript $a$. 
Let $\hat{\mathcal{S}}_Q$ be a $2^Q\times 2^Q$ matrix defined by $\hat{\mathcal{S}}_{Q,ij}=\delta_{i,j+1}+\delta_{i,j-1}$. 
Then, up to an overall factor, $\hat{p}^2=2-\hat{\mathcal{S}}_Q$.
We have
\begin{align}
    \hat{\mathcal{S}}_1=\hat{\sigma}_x, 
\end{align}
\begin{align}
    \hat{\mathcal{S}}_2=
    \left(
    \begin{array}{cc}
    \hat{\mathcal{S}}_1 & \hat{\Sigma}_-\\
    \hat{\Sigma}_+ & \hat{\mathcal{S}}_1
    \end{array}
    \right), 
\end{align}
where
$\hat{\Sigma}_\pm=\frac{\hat{\sigma}_x\pm \mathrm{i}\hat{\sigma}_y}{2}$, 
and more generally, 
\begin{align}
    \hat{\mathcal{S}}_{Q+1}=
    \left(
    \begin{array}{cc}
    \hat{\mathcal{S}}_Q & \hat{\Sigma}_-^{\otimes Q}\\
    \hat{\Sigma}_+^{\otimes Q} & \hat{\mathcal{S}}_Q
    \end{array}
    \right).  
\end{align} 
We can show that $\mathcal{S}_Q$ is a sum of $2^Q-1$ Pauli strings. 
There are $2^{l-1}$ Pauli strings of length $l$. 
To see this, note that
\begin{align}
    \hat{\mathcal{S}}_{Q+1}=
    \hat{\mathcal{S}}_Q\otimes\textbf{1}_2
    +
    \hat{\Sigma}_-^{\otimes Q}\otimes\hat{\Sigma}_+
    +
    \hat{\Sigma}_+^{\otimes Q}\otimes\hat{\Sigma}_-. 
\end{align}

$\Sigma_-^{\otimes Q}\otimes\Sigma_+$ and $\Sigma_+^{\otimes Q}\otimes\Sigma_-$ are written as the sum of $Q+1$ $\hat{\sigma}_x$ or $\hat{\sigma}_y$. 
They are related by $\hat{\sigma}_y\leftrightarrow -\hat{\sigma}_y$, so the terms with odd numbers of $\hat{\sigma}_y$ cancel when we take the sum of $\Sigma_-^{\otimes Q}\otimes\Sigma_+$ and $\Sigma_+^{\otimes Q}\otimes\Sigma_-$. 
The general form is 
\begin{align}
    \hat{\mathcal{S}}_Q
    =
    \sum_l
    \frac{1}{2^{l-1}}
    \left(\sum
    {\rm length}\ l\ {\rm Pauli\ strings}
    \right)
    \otimes\textbf{1}_{2^{Q-l}},
\end{align}
where the sum over Pauli strings is taken only for the ones containing only $\hat{\sigma}_x$ and $\hat{\sigma}_y$, with an even number of $\hat{\sigma}_y$'s. 
There are $2^{l-1}$ such Pauli strings. 
The number of Pauli strings needed is $\sum_{l=1}^Q 2^{l-1} = 2^Q-1$. 
Explicit expressions are 
\begin{align}
    \hat{\mathcal{S}}_1=\hat{\sigma}_x\, 
\end{align}
for $Q=1$, 
\begin{align}
    \hat{\mathcal{S}}_2
    =
    \hat{\mathcal{S}}_1\otimes\textbf{1}_2
    +
    \frac{\hat{\sigma}_x\otimes\hat{\sigma}_x+\hat{\sigma}_y\otimes\hat{\sigma}_y}{2}\, . 
\end{align}
for $Q=2$, 
\begin{align}
    \hat{\mathcal{S}}_3
    &=
    \hat{\mathcal{S}}_2\otimes\textbf{1}_2
    +
    \frac{\hat{\sigma}_x\otimes\hat{\sigma}_x\otimes\hat{\sigma}_x-\hat{\sigma}_x\otimes\hat{\sigma}_y\otimes\hat{\sigma}_y+\hat{\sigma}_y\otimes\hat{\sigma}_x\otimes\hat{\sigma}_y+\hat{\sigma}_y\otimes\hat{\sigma}_y\otimes\hat{\sigma}_x}{4}
\end{align}
for $Q=3$, 
\begin{align}
    \hat{\mathcal{S}}_4
    &=\hat{\mathcal{S}}_3\otimes\textbf{1}_2
    \nonumber\\
    &
    +
    \frac{1}{8}\left\{
    \hat{\sigma}_x\otimes\hat{\sigma}_x\otimes\hat{\sigma}_x\otimes\hat{\sigma}_x
    +
    \hat{\sigma}_x\otimes\hat{\sigma}_x\otimes\hat{\sigma}_y\otimes\hat{\sigma}_y
    +
    \hat{\sigma}_x\otimes\hat{\sigma}_y\otimes\hat{\sigma}_x\otimes\hat{\sigma}_y
    -
    \hat{\sigma}_x\otimes\hat{\sigma}_y\otimes\hat{\sigma}_y\otimes\hat{\sigma}_x
    \right.
    \nonumber\\
    &
    \quad\ \ 
    \left.
    +
    \hat{\sigma}_y\otimes\hat{\sigma}_x\otimes\hat{\sigma}_x\otimes\hat{\sigma}_y
    -
    \hat{\sigma}_y\otimes\hat{\sigma}_x\otimes\hat{\sigma}_y\otimes\hat{\sigma}_x
    -
    \hat{\sigma}_y\otimes\hat{\sigma}_y\otimes\hat{\sigma}_x\otimes\hat{\sigma}_x
    -
    \hat{\sigma}_y\otimes\hat{\sigma}_y\otimes\hat{\sigma}_y\otimes\hat{\sigma}_y
    \right\}
    \nonumber\\
\end{align}
for $Q=4$, and
\begin{align}
    \hat{\mathcal{S}}_5
    &=
    \hat{\mathcal{S}}_4\otimes\textbf{1}_2
    \nonumber\\
    &
    +
    \frac{1}{16}\Bigl(
    \hat{\sigma}_x \otimes  \hat{\sigma}_x \otimes  \hat{\sigma}_x \otimes  \hat{\sigma}_x\otimes  \hat{\sigma}_x 
    \nonumber\\
    &\qquad
    -
    \hat{\sigma}_y \otimes  \hat{\sigma}_y \otimes  \hat{\sigma}_x \otimes  \hat{\sigma}_x\otimes  \hat{\sigma}_x
    - 
    \hat{\sigma}_y \otimes  \hat{\sigma}_x \otimes  \hat{\sigma}_y \otimes  \hat{\sigma}_x\otimes  \hat{\sigma}_x
    -
    \hat{\sigma}_y \otimes  \hat{\sigma}_x \otimes  \hat{\sigma}_x \otimes  \hat{\sigma}_y\otimes  \hat{\sigma}_x
    \nonumber\\
    &\qquad
    +
    \hat{\sigma}_y \otimes  \hat{\sigma}_x \otimes  \hat{\sigma}_x \otimes  \hat{\sigma}_x\otimes  \hat{\sigma}_y
    - 
    \hat{\sigma}_x \otimes  \hat{\sigma}_y \otimes  \hat{\sigma}_y \otimes  \hat{\sigma}_x\otimes  \hat{\sigma}_x
    - 
    \hat{\sigma}_x \otimes  \hat{\sigma}_y \otimes  \hat{\sigma}_x \otimes  \hat{\sigma}_y\otimes  \hat{\sigma}_x
    \nonumber\\
    &\qquad
    + 
    \hat{\sigma}_x \otimes  \hat{\sigma}_y \otimes  \hat{\sigma}_x \otimes  \hat{\sigma}_x\otimes  \hat{\sigma}_y 
    -
    \hat{\sigma}_x \otimes  \hat{\sigma}_x \otimes  \hat{\sigma}_y \otimes  \hat{\sigma}_y\otimes  \hat{\sigma}_x
    + 
    \hat{\sigma}_x \otimes  \hat{\sigma}_x \otimes  \hat{\sigma}_y \otimes  \hat{\sigma}_x\otimes  \hat{\sigma}_y
    \nonumber\\
    &\qquad
    + 
    \hat{\sigma}_x \otimes  \hat{\sigma}_x \otimes  \hat{\sigma}_x \otimes  \hat{\sigma}_y\otimes  \hat{\sigma}_y
    \nonumber\\
    &\qquad
    - 
    \hat{\sigma}_x \otimes  \hat{\sigma}_y \otimes  \hat{\sigma}_y \otimes  \hat{\sigma}_y\otimes  \hat{\sigma}_y
    - 
    \hat{\sigma}_y \otimes  \hat{\sigma}_x \otimes  \hat{\sigma}_y \otimes  \hat{\sigma}_y\otimes  \hat{\sigma}_y
    - 
    \hat{\sigma}_y \otimes  \hat{\sigma}_y \otimes  \hat{\sigma}_x \otimes  \hat{\sigma}_y\otimes  \hat{\sigma}_y 
    \nonumber\\
    &\qquad
    - 
    \hat{\sigma}_y \otimes  \hat{\sigma}_y \otimes  \hat{\sigma}_y \otimes  \hat{\sigma}_x\otimes  \hat{\sigma}_y 
    + 
    \hat{\sigma}_y \otimes  \hat{\sigma}_y \otimes  \hat{\sigma}_y \otimes  \hat{\sigma}_y\otimes  \hat{\sigma}_x 
    \Bigl)
\end{align}
for $Q=5$ \footnote{The expression for Q=5 was obtained using the Pauli string decomposer algorithm proposed in Appendix D of \cite{Mendicelli:2023eel}.}. 

For the operator $\hat{p}^2$ defined by \eqref{p^2-coordinate-basis} in this simple system, we could improve the circuit so that exponentially many Pauli strings can be expressed by a smaller number of gates, for example by using the efficient block encoding of Ref.~\cite{ty2024doublelogarithmic}. 
However, the situation becomes complicated when we consider more nontrivial examples such as the Kogut-Susskind Hamiltonian, in which we cannot find such a simple alternative method without a lot of new research effort.
\subsection{Problem with Fock basis}\label{sec:naive_Fock_basis}
Next, let us consider the truncation in the Fock basis. 
Again, we focus on one boson and drop the subscript $a$. 
Let us define the creation and annihilation operators as 
\begin{align}
    \hat{A}^\dagger
    =
    \sqrt{m\omega}\hat{x}
    -
    \frac{\mathrm{i}\hat{p}}{\sqrt{2m\omega}}\, , 
    \qquad
    \hat{A}
    =
    \sqrt{m\omega}\hat{x}
    +
    \frac{\mathrm{i}\hat{p}}{\sqrt{2m\omega}}\, . 
\end{align}
Here, $m$ and $\omega$ are parameters that characterize the choice of basis. 
The Fock vacuum $\ket{0}$ is given by $\hat{A}\ket{0}=0$. 
We can obtain other Fock states as 
\begin{align}
    \ket{j}
    =
    \frac{(A^\dagger)^j}{\sqrt{j!}}\ket{0}\, . 
\end{align}
We introduce a truncation $j<\Lambda=2^Q$ and regularize the Fock space by introducing the cutoff $\Lambda$ for the excited modes of the harmonic oscillators. 
For each harmonic oscillator, we assign $Q=\log_2\Lambda$ qubits. 
We use the compact mapping
\begin{align}
\ket{j}  = \ket{ b_0} \ket{ b_1} \cdots\ket{b_{Q-1}}
\end{align}
for the energy level $j=0,1,\cdots,\Lambda-1$, where we use the binary decomposition
\begin{align}
j = b_0 + 2b_1+\cdots+2^{Q - 1}b_{Q - 1}\, .  
\end{align}
One can map the creation operator $A^\dagger$ to 
\begin{align}
\hat{A}^\dagger = \sum\limits_{j = 0}^{\Lambda - 2} {\sqrt {j + 1} } |j + 1\rangle \langle j|. 
\label{eq:A_dagger}
\end{align}

The truncated version of $\hat{x}$ and $\hat{p}$ are obtained from the truncated version of $\hat{A}^\dagger$ and $\hat{A}$, as 
\begin{equation}
\hat{x} 
\propto
\hat{A}^\dagger + \hat{A}
=
\left(
\begin{array}{ccccc}
0 & \sqrt{1} & 0 &\cdots &0\\
\sqrt{1} & 0 & \sqrt{2} &\cdots &0  \\
0 & \sqrt{2} & \ddots & \ddots & \vdots  \\
\vdots & \vdots & \ddots & 0 & \sqrt{\Lambda -1} \\
0 & 0 & \cdots & \sqrt{\Lambda -1} & 0 \\
\end{array}
\right)
\end{equation}
and
\begin{equation}
\hat{p} 
\propto
\mathrm{i}\left(\hat{A}^\dagger-\hat{A}\right)
=
\mathrm{i}
\left(
\begin{array}{ccccc}
0 & -\sqrt{1} & 0 &\cdots &0\\
\sqrt{1} & 0 & -\sqrt{2} &\cdots &0  \\
0 & \sqrt{2} & \ddots & \ddots &\vdots  \\
\vdots & \vdots & \ddots & 0 & -\sqrt{\Lambda -1} \\
0 & 0 & \cdots & \sqrt{\Lambda -1} & 0 \\
\end{array}
\right)\, . 
\end{equation}

The similarity to $\hat{\mathcal{S}}$ in the coordinate basis \eqref{p^2-coordinate-basis} is clear. 
We counted the number of Pauli strings needed to express these operators for $Q=(1,2,\cdots,14)$ using Qiskit (which implements the algorithm proposed by Ref.~\cite{Hantzko:2023aal}) and observed that the number is $Q\cdot 2^{Q-1}$ for all the cases we studied; see Table~\ref{table:x_p_Fock_basis}. 
Therefore, we conjecture that this is the analytic expression valid for any $Q$.\\

\begin{table}
{\footnotesize
\begin{tabular}{|c||cccccccccccccc|}
\hline
$Q$ & 1 & 2 & 3 & 4 & 5 & 6 & 7 & 8 & 9 & 10 & 11 & 12 & 13 & 14\\
\hline
\hline
$\Lambda=2^Q$ & 2 & 4 & 8 & 16 & 32 & 64 & 128 & 256 & 521 & 1024 & 2048 & 4096 & 8192 & 16384\\
\hline
\# Pauli & 
1 & 4 & 12 & 32 & 80 & 192 & 448 & 1024 & 2304 & 5120 & 11264 & 24576 & 53248 & 114688\\
\hline
\end{tabular}
}
\caption{The number of Pauli strings needed to encode $\hat{x}$ or $\hat{p}$ in the Fock basis, for the truncation level $\Lambda=2^Q$ represented by $Q$ qubits.
We can see that the number of Pauli strings is $Q\cdot 2^{Q-1}$. 
}\label{table:x_p_Fock_basis}
\end{table}

\begin{figure}[ht]
    \centering
    \begin{subfigure}{0.45\textwidth}
        \centering
        \includegraphics[width=\linewidth]{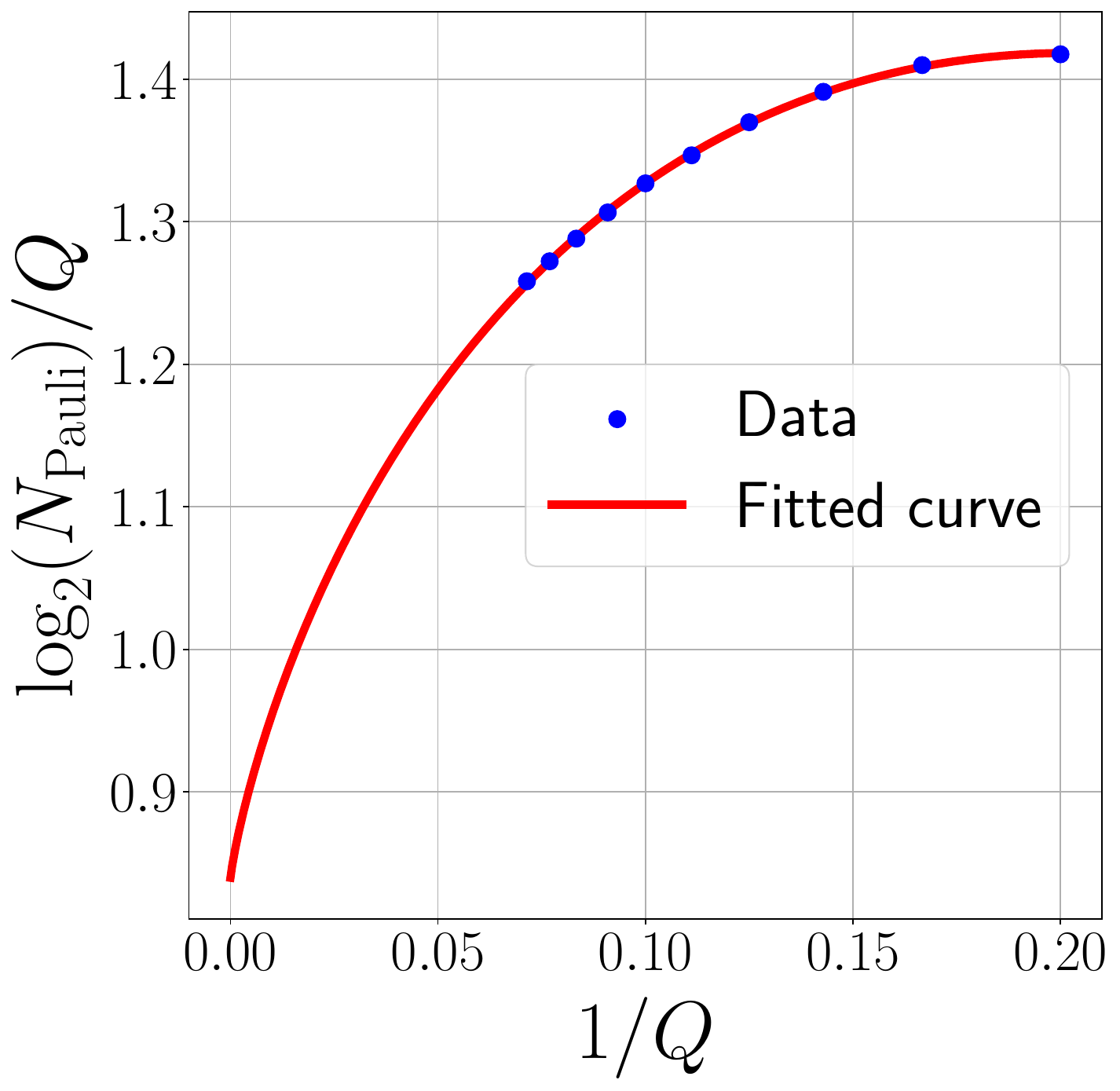}
        {Double well potential $(m\hat{x} +g\hat{x}^2+g\mu^2)^2$, where $m=g=\mu=1$. }
        \label{fig:sub2__}
    \end{subfigure}
    \hfill
    \begin{subfigure}{0.45\textwidth}
        \centering
    \includegraphics[width=\linewidth]{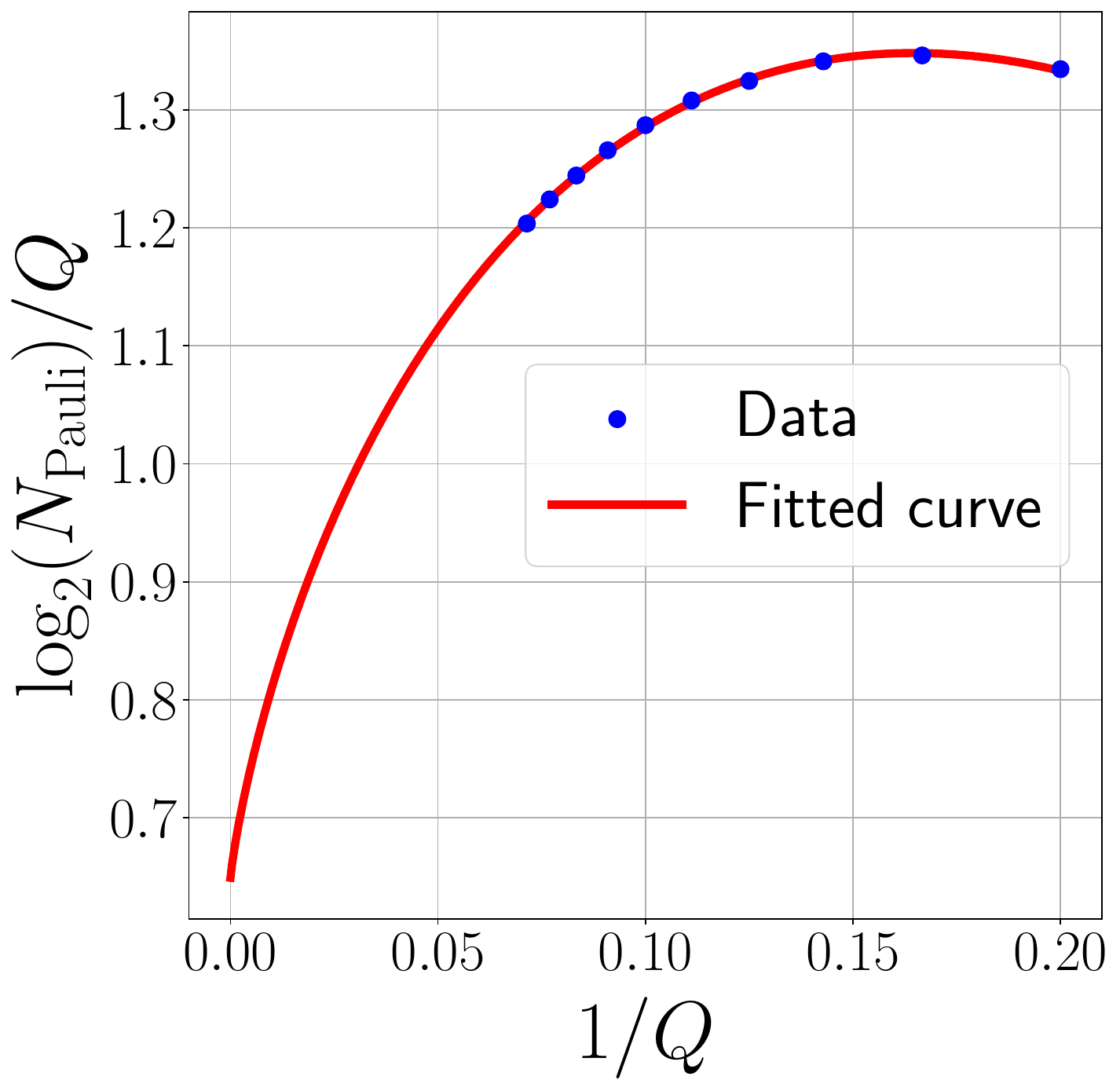}
    {Anharmonic oscillator, $(m\hat{x} + g \hat{x}^3)^2$, where $m=g=1$. }
        \label{fig:sub3__}
    \end{subfigure}
    \caption{Fits of the number of Pauli strings $N_{\rm Pauli}$ from Table~1 of \cite{Mendicelli:2024ryt} obtained by extending Table~1 of Ref.~\cite{Mendicelli:2024ryt} up to 14 qubits. 
    The fit ansatz is $\frac{1}{Q}\log N_{\rm Pauli}=a+\frac{b+c\log Q}{Q}$. 
    We can see a nonzero value of $a$ that indicates the exponential growth of $N_{\rm Pauli} $ concerning the number of qubits $Q$.} \label{fig:SQM_Pauli_fit}
\end{figure}

There is no reason to expect hypothetical cancellations of Pauli strings $N_{\rm Pauli}$ in an interacting Hamiltonian. 
As a concrete example, in Fig.~\ref{fig:SQM_Pauli_fit}, we show the scaling of the number of Pauli strings in the bosonic potential term in supersymmetric quantum mechanics studied in Ref.~\cite{Mendicelli:2024ryt}. 
$\frac{1}{Q}\log N_{\rm Pauli}$ converges to a nonzero value as $Q\to\infty$, demonstrating the exponential growth of $N_{\rm Pauli}$ as function of $Q$.

Again, in this simple example, we could improve the circuit for $\hat{A}^\dagger$ defined by \eqref{eq:A_dagger}; see Ref.~\cite{Simon:2025pbo} for recent new efficient algorithms for block encoding. 
However, the situation becomes complicated when we consider more nontrivial examples such as the Kogut-Susskind Hamiltonian.
\subsection{Cost of programming}\label{sec:Pauli_string_expansion_is_hard}
When we use Pauli strings to build circuits, the exponential growth of the number of Pauli strings is problematic for obvious reasons. 
Among them, the most serious ones are (i) the circuit becomes exponentially long, and (ii) it has exponentially large classical input, which becomes out of control quickly --- we would not be able to store the coefficients in classical memory.

Even if the growth of the number of Pauli strings is not too fast, if we cannot determine the coefficients in the Pauli string expansion by hand, we need to determine them numerically \textit{on a classical computer}. 
Unless we know certain underlying patterns in the Hamiltonian, the cost increases exponentially with the number of qubits. 
For $n$ qubits, the worst-case scenario is that we have to check all $4^n$ possible Pauli strings; see, e.g.~\cite{Hantzko:2023aal}.
This computation has to be done only once before starting the quantum simulation; however, this is already a challenging task that destroys the potential quantum advantage. 

While Pauli string expansions are commonly used in fermionic systems due to fermion-to-qubit mappings such as the Jordan--Wigner and Bravyi--Kitaev transformations,
other types of operator expansions may be more appropriate in more general settings, such as bosonic systems.
In more general cases like bosonic systems, 
one may consider different expansions.
For instance, in the case of $d$-sparse Hamiltonian, where the number of nonzero entries in each row and column is bounded by $d$,
it is possible to decomposed the original Hamiltonian into a $O(d^2)$  1-sparse Hamiltonians~\cite{AharonovTaShma,Berry2013,Berry2007}.
This decomposition is based on a graph coloring algorithm applied to a bipartite graph representation of the Hamiltonian.
The input size of the coloring problem is $O(2^n)$, corresponding to the Hilbert space dimension.
Therefore, even if the classical algorithm solves the coloring problem in time polynomial in the input size, the overall complexity is still exponential in the number of qubits $n$.

There are various works that develop efficient quantum simulation algorithms for sparse Hamiltonians; for instance, see~\cite{AharonovTaShma,Berry2013,Berry2007,Childs2011,Wiebe2011,Berry2015,Kieferova2019,Low2019}.
These quantum algorithms typically rely on oracles to efficiently encode the structure of a sparse matrix $A$. Specifically, in the case of~ \cite{Gilyen2018}:
The first oracle (denoted $O_r$) tells us which columns are nonzero in a given row. 
The second oracle (denoted $O_c$) does the same, but for columns.
The third oracle (denoted $O_A$) returns the actual value of the matrix element $A_{ij}$ for a given row $i$ and column $j$.

While considerable effort has been devoted to reducing the query complexity of these oracles, this does not directly translate into end-to-end efficiency for quantum simulations, nor does it address how to construct the full circuit required to realize such simulations efficiently.
This is because so-called ``efficient" simulations often assume the existence of idealized oracles, whereas designing these oracles explicitly in terms of native quantum gates can be highly nontrivial and resource-intensive.

Recent works~\cite{Camps:2022uyk, Camps2023} explore explicit circuit constructions for block encodings of sparse matrices, and other efforts aim to automate the process beyond simple sparsity patterns~\cite{Kuklinski:2024idb}. 
Let us take the oracles for sparse operators as an illustrative example. 
A fundamental assumption, often not satisfied in physically relevant cases, is that the sparse operator is explicitly known analytically. 
In practice, however, determining the full structure of a sparse operator may itself require difficult classical computations. 
If we attempt to identify all nonzero entries one by one, the cost can become exponential, as the total number of elements in the Hamiltonian scales exponentially with the number of qubits.

Even when all the classical processing --- such as identifying the nonzero elements and decomposing the Hamiltonian into 1-sparse components --- is completed, constructing the corresponding oracles using native quantum gates remains nontrivial.
For example, Ref.~\cite{Camps2023} demonstrated how to explicitly construct such oracles in specific cases. 
However, extending these methods to more general settings, such as the Kogut–Susskind Hamiltonian, remains an open problem.

There is no general way to bypass this issue unless simple analytic formulas are available, as in certain cases like scalar quantum field theory~\cite{Jordan:2012xnu} or the orbifold lattice Hamiltonian~\cite{Buser:2020cvn,Bergner:2024qjl,Halimeh:2024bth}, where the structure of the Hamiltonian is known a priori.

Perhaps it is worth mentioning the Harrow--Hassidim--Lloyd (HHL) algorithm~\cite{HHL2009}, a well-known quantum algorithm for solving systems of linear equations. 
While HHL provides an exponential speedup in principle, its practical utility depends on satisfying several stringent conditions. 
For instance, the input matrix must be sparse and well-conditioned, and its entries must be accessible via efficient oracles. Moreover, preparing the input state and extracting the solution (i.e., avoiding classical upload/download bottlenecks) must also be implemented efficiently. 
These requirements highlight a broader challenge in quantum algorithms: the theoretical efficiency often assumes idealized conditions --- such as oracle access and efficient encoding --- that may be hard to achieve in realistic settings. 
These limitations, which are widely recognized in the community, serve as a clear example of how the gap between query complexity and full-circuit realizability can significantly impact the efficiency of quantum algorithms.

\section{When does it matter?}\label{sec:when_does_it_matter}
The exponential growth of the number of gates with $Q$ may or may not be a serious problem, depending on how large $Q$ is needed for practical applications. 
A common view among the physics research community is the following.
When all parameters of the model, other than the truncation level $\Lambda=2^Q$, are fixed, the truncation effect on the low-energy states is expected to decay exponentially with $\Lambda$ as $\epsilon\sim e^{-c\Lambda}$, with some constant $c$~\cite{Klco:2018zqz,Tong:2021rfv,Hanada:2022pps}. 
Then, to get a truncation error that is sufficiently small, one does not have to take $\Lambda$ too large ($\Lambda\sim\log\left(\frac{1}{\epsilon}\right)$), and hence, the power growth with $\Lambda$ is not a serious problem.
Numerical tests in a U(1) gauge theory in 1 dimension~\cite{Ciavarella:2025tdl} have recently shown that an upper bound on $\Lambda$ can be computed and it decays even faster than exponentially at fixed parameters of the model.

The crucial assumption in the above is that all model parameters other than the truncation level are fixed. 
This is, of course, not always a valid assumption, and we want to explicitly show that a large value of $\Lambda$ is needed with a few examples below.

\subsection{Physical examples}\label{sec:physical_examples}

As a simple example, let us imagine particles in a three-dimensional box of volume $V=L^3$ and that we are interested in processes that require microscopic interaction at a fixed distance scale. 
If we want to reach the thermodynamic limit where $V$ is very large, but do so at fixed particle density, we would need to assign $\Lambda\gtrsim L$ points to the $x$, $y$, and $z$ coordinates of each particle to maintain fixed spatial resolution.\\
\\
Another situation that requires large values of $\Lambda$ is the large-$N$ limit of matrix models and Yang-Mills theory. 
These models are particularly interesting in the context of the holographic description of black holes~\cite{Maldacena:2023acv}.
As a concrete setup, let us consider the motion of a probe (e.g., D-brane) moving in a black hole background. 
In this case we need to be able to resolve the position within a variance of order $N^0$ and with typical values of order $\sqrt{N}$~\cite{Hanada:2021ipb}. 
Therefore, $\Lambda\gtrsim\sqrt{N}$ is a reasonable choice, and we have to take the large-$N$ limit.\\
\\
Yet another important example is quantum field theory, most notably Yang-Mills theory, which is the underlying theoretical framework of QCD.
To study quantum field theory on a quantum computer, we first introduce a spatial lattice in a finite spatial volume and with a finite lattice spacing $a$ (this must be distinguished from the spacing associated with the truncation of the boson's Hilbert space, $\delta_x$), so that the number of bosons $B$ becomes finite.
Then, we truncate the Hilbert space of each boson introducing $\Lambda$, representing the largest values a specific bosonic operator can have on a state in the local Hilbert space. 
To recover the original theory, we first remove the truncation ($\Lambda \to \infty$, $\delta_x \to 0$, $\Lambda \cdot \delta_x \to \infty$) and then send the lattice spacing to zero ($a \to 0$), keeping the physical volume sufficiently large. 
Many technical complications can arise associated with ultraviolet divergences. 
In lattice gauge theory, the unitary link variable $U_{\mu,\vec{\mathrm{x}}}$ is related to the gauge field $A_{\mu,\vec{\mathrm{x}}}$ by $U_{\mu,\vec{\mathrm{x}}}=\exp\left(\mathrm{i}agA_{\mu,\vec{\mathrm{x}}}\right)$, where $\mu$ labels the spatial dimensions and $\vec{\mathrm{x}}$ is the spatial coordinate that should not be confused with $\vec{x}$ in the previous sections; the gauge field $A_{\mu,\vec{\mathrm{x}}}$ is the counterpart of the coordinate $\vec{x}$. 
For $(2+1)$- and $(3+1)$-dimensional theories, typical values of $A_{\mu,\vec{\mathrm{x}}}$ increase as $a \to 0$. 
Due to the asymptotic freedom, lattice perturbation theory (see e.g.,~Ref.~\cite{Lepage:1992xa}) becomes valid at short distance scales. 
From the scaling of the expectation value of the plaquette, we see that the typical value of field strength $F_{\mu\nu,\vec{\mathrm{x}}}$ in the ground state scales as $\frac{1}{a^2}$, which in turn means that the size of the ground-state wave function in terms of gauge field $A_{\mu,\vec{\mathrm{x}}}$ spreads as $\frac{1}{a}$. 
To cover most of the wave function while keeping the resolution fixed, $\Lambda\gtrsim\frac{1}{a}$ is required.\\
\\
In the above examples, the exponential increase with $Q$ corresponds to the increase in power law with the system size, such as the size of $L$, $N$, or $\frac{1}{a}$ (the latter is proportional to the number of lattice points in each dimension for fixed physical volume).
It is well known in the Kogut-Susskind Hamiltonian research literature that quantum resources scale polynomially with the system size.
For this reason, it may be better to say that the use of the quantum Fourier transform that we advocated before changes the resource growth rate from polynomial to logarithmic, rather than saying that the other approaches suffer from an exponential slowdown.

The quantum Fourier transform is one of the most important subroutines of quantum algorithms which will surely be implemented in a useful fault-tolerant digital computer.
In fact, Ref.~\cite{Mayer:2024yfm} has already experimentally benchmarked the quantum Fourier transform primitive on three logical qubits. 
Whenever possible, we should use the universal framework that utilizes the quantum Fourier transform, due to, not only its simplicity in translating algorithms to circuits, but also its demonstrated asymptotic speedup in $Q$. 
However, what if the Hamiltonian does not take the form of Eq.~\eqref{generic_Hamiltonian}?  
The Kogut-Susskind Hamiltonian~\cite{Kogut:1974ag}, which has been a popular option for Yang-Mills theory and QCD, is different from~\eqref{generic_Hamiltonian} due to its use of compact variables. 
For generic gauge groups such as SU(3), there is no known way to introduce the truncated coordinate (magnetic) and momentum (electric) bases and connect them with a Fourier transform.
Irreducible representations play an analogous role to plane waves due to the Peter-Weyl theorem, which makes the use of the momentum basis technically complicated but not impossible~\cite{Byrnes:2005qx}. 
It is difficult to define a simple truncation in the coordinate basis, because there is no obvious way to discretize the group manifold. 
Refs.~\cite{Garofalo:2023zkd,Romiti:2023hbd,Jakobs:2025rvz} show attempts to build the coordinate basis for SU(2) Yang-Mills theory.  
Overall, so far, the Kogut-Susskind formulation does not enjoy the speedup we have seen applied to the universal framework.
Strictly speaking, the truncated Kogut-Susskind Hamiltonian is very complicated already for the SU(3) gauge group, and it is hard to even write it down explicitly~\cite{Balaji:2025afl}.
On the other hand, the orbifold lattice formulation~\cite{Buser:2020cvn,Bergner:2024qjl,Kaplan:2002wv} has the form in Eq.~\eqref{generic_Hamiltonian} and enjoys the speed-up of the universal framework. 

Note also that, unless one finds an efficient way to determine the Pauli string expansion of the Kogut-Susskind Hamiltonian, the problem explained in Sec.~\ref{sec:Pauli_string_expansion_is_hard} introduces further difficulty. 
A SU($N$) link variable is not just a simple product of $N^2-1$ bosons. 
Rather, $N^2-1$ bosons are intricately intertwined, leading to a severe scaling $4^n$ for the cost of Pauli string expansion we have seen in Sec.~\ref{sec:Pauli_string_expansion_is_hard}. Intuitively, we expect $n=Q(N^2-1)$ and $4^n=4^{Q(N^2-1)}$, although it is not easy to quantify it precisely. 
For $(3+1)$-dimensional theory, combined with $\Lambda=2^Q\sim\frac{1}{a}$, the scaling is roughly $4^{Q(N^2-1)}\sim a^{-2(N^2-1)}$, which seems challenging already for $N=3$. 
For plaquette terms, we need to take a tensor product of four link variables, which would lead to $4^{4Q(N^2-1)}\sim a^{-8(N^2-1)}$. 

\subsection{Comparison to previous works}\label{sec:comparison}
Ref.~\cite{Rhodes:2024zbr} introduced an efficient simulation protocol for a SU($N$) lattice gauge theory assuming sparse oracles. 
Essentially, the problem reduces to an explicit circuit realization of the sparse oracle describing the unitary link variables in terms of native gates.
This is a highly nontrivial task, except for simple gauge groups, and needs to be worked out for each gauge group separately.
Because of the Peter-Weyl theorem, on the electric basis, each link variable is given by the irreducible representations of SU($N$). 
For a state in an irreducible representation $\mathrm{r}$, the action of the link variable multiplies the fundamental representation $\mathrm{f}$, and we will need a decomposition of $\mathrm{f}\otimes\mathrm{r}$ in terms of irreducible representations, schematically, 
\begin{align}
\ket{\mathrm{f}\otimes\mathrm{r}}
=
\sum_{\mathrm{r}'}
\ket{\mathrm{r}'}
\bra{\mathrm{r}'}\ket{\mathrm{f}\otimes\mathrm{r}}\, , 
\end{align}
where the overlap $\bra{\mathrm{r}'}\ket{\mathrm{f}\otimes\mathrm{r}}$ is the Clebsch-Gordan coefficient.
Written more explicitly, 
\begin{align}
R^{\rm [f]}_{ij}(U)R^{\rm [r]}_{kl}(U)
=
\sum_{{\rm r'},I,J}
C_{{\rm r'}IJ; {\rm f}ij; {\rm r}kl}
R^{\rm [r']}_{IJ}(U)\, . 
\end{align}
The dimension of the representations $\mathrm{r}$, $\mathrm{r}'$ we need to handle increases exponentially with the number of qubits $Q$. 
For example, in the case of SU(2), irreducible representations are described by spin $j=0,\frac{1}{2}, 1, \frac{3}{2},\cdots$. 
The dimension of the spin-$j$ representation is $2j+1$. 
Therefore, the maximum spin in the truncated Hilbert space $j_{\rm max}$ is given by 
\begin{align}
    2^Q
    =
    1+2+3+\cdots+(2j_{\rm max}+1)
    =
    (j_{\rm max}+1)(2j_{\rm max}+1)\, ,  
\end{align}
and the asymptotic scaling is 
\begin{align}
    j_{\rm max} \simeq 2^{(Q-1)/2}\, . 
\end{align}
Note that the growth is much faster for SU($N$) with  $N\ge 3$.
Furthermore, the number of nonzero Clebsch-Gordan coefficients increases compared to SU(2); see Ref.~\cite{Balaji:2025afl} for a concrete analysis. 
This has been studied extensively by the community, and it is clear that even putting the representation in a memory on a classical device is exponentially difficult with regard to $Q$, unless a clear pattern is identified analytically or $Q$ is a very small number.
The oracle-based approach of Ref.~\cite{Rhodes:2024zbr} is a great tool to provide resource estimations, but it becomes impractical for a concrete realization of the Hamiltonian time evolution unless a viable implementation of the oracles in terms of native gates is found.
As a compilation problem, this would remain classically hard. 
Note also that, if one tries to program the circuit on a quantum computer without using efficient oracles, one would have to upload very large classical input (e.g., all Clebsch-Gordan coefficients) into a quantum device, which is a highly nontrivial process. 
The size of classical input would be the same as the above estimate in the worst case: $4^{Q(N^2-1)}\sim a^{-2(N^2-1)}$ for each link and $4^{4Q(N^2-1)}\sim a^{-8(N^2-1)}$ for each plaquette.
Note that the simple cases discussed in Sec.~\ref{sec:naive_coord_basis} and Sec.~\ref{sec:naive_Fock_basis} can be considered as instances where constructing oracles was relatively straightforward.

A practical approach that may mitigate this problem is to find an improved Hamiltonian that does not require a very small lattice spacing $a$, so that one can keep the truncation parameter $\Lambda$ relatively small while achieving a good approximation to the continuum limit.\footnote{
Note, however, there is a trade off: if more link variables are introduced to the interaction terms associated with the improvement, then one needs to deal with tensor products of many link variables that pushes up the computational complexity at fixed $\Lambda$. 
} 
Note that the effect of finite lattice spacing can be studied using the Euclidean path integral on classical computers~\cite{Hanada:2022pps}. 
Another practical and simple approach is to regard the unitary link variable as the infinite scalar-mass limit of the complex link variable, even at the regularized level. 
If one takes this route, as a bonus, one can use the universal framework. 

In Appendix~\ref{appendix:cost_analyses}, we discuss the cost of quantum simulation based on several known approaches. In all known cases that we discuss, we find no systematic analysis in the literature on the cost of simulations as the truncation level $\Lambda$ increases.
This is because the Hamiltonian in these approaches is too complicated and does not easily allow people to estimate the needed quantum resources. 
Unless someone achieves a technical breakthrough for those approaches known in the literature, we do not see how to solve their exponential inefficiency.

\subsection{Gauge invariance}\label{sec:gauge_invariance}
The use of gauge-invariant Hilbert space causes the same problem, even for matrix models or orbifold lattices, and the consequences can be much more dramatic. 
Due to the non-local nature of the gauge-invariant Hilbert space, we cannot use the quantum Fourier transform. 
Furthermore, the cost of determining the Pauli string expansion increases proportionally to $(\mathrm{dim}\mathcal{H})^2$, unless one finds a systematic way to find the coefficients without relying on brute-force computations. 
At low energy, one may appeal to the sparsity of the Hamiltonian to find an efficient simulation protocol, but that would still require the explicit determination of all the necessary non-zero matrix elements, which may be exponentially hard on classical computers.
Therefore, merely designing a circuit can already be exponentially hard, not only with regard to $Q$ but also $B$. 
Then, practically, the circuit can be designed only when the quantum simulation can be emulated on classical computers.\footnote{This problem can be avoided by using the energy eigenbasis. However, if one could find the energy eigenstates, one would not need to use a quantum computer.}
Still, the gauge-invariant Hilbert space can be advantageous for \textit{classical simulations} due to the reduction of the dimension of the Hilbert space that enables one to diagonalize the Hamiltonian for small lattices truncated at low energy.
We also note that the gauge symmetry is important because it enables us to introduce massless vector field compatible with Lorentz symmetry without having negative-norm states. In lattice formulations, even when the truncation in the Hilbert space is removed, Lorentz symmetry is broken due to a finite lattice spacing. 
Adherence to the singlet Hilbert space can make it practically difficult (e.g. in terms of resources) to go to sufficiently small lattice spacing and hence can lead to larger breaking of Lorentz symmetry.

\section{Discussions}\label{sec:discussion}
In this paper, we pointed out the advantage of a simple truncation and simulation scheme for bosons that admits a straightforward application of the quantum Fourier transform, applicable to a rather general class of bosonic Hamiltonians that takes the form \eqref{generic_Hamiltonian}. 
When the number of qubits assigned to each boson needs to increase with the system size, the quantum Fourier transform leads to a significant speedup. 
In Ref.~\cite{Halimeh:2024bth}, three of the authors (M.~H., S.~M., and E.~R.) and their collaborators emphasized the value of a simple, universal approach that does not rely on the specifics of the systems involved. 
The significant speedup we highlight in this study further supports this perspective: the efficiency gained is a natural outcome of our efforts to simplify and generalize the setup, rather than introducing more complex structures that may become too specific to certain physical systems. 
This brings us to an important and potentially contentious question regarding the study of non-Abelian lattice gauge theory: are the prevalent approaches -- such as those based on the Kogut-Susskind Hamiltonian and/or gauge-invariant Hilbert space formulations -- effective for quantum simulations?

It is important to note that there is nothing inherently wrong with these approaches as theoretical frameworks. 
They have led to many important analytical results. 
However, when applied to quantum simulations, they may not offer a straightforward path forward. 
The Kogut-Susskind Hamiltonian, for instance, involves compact variables, and it is already challenging to express the truncated Hamiltonian explicitly, and classical processing can be a bottleneck. 
Although such complexity might be justifiable if it led to more efficient simulations, our findings suggest that, in reality, it impedes efficiency rather than enhancing it. 
Sociologically, the preference for the Kogut-Susskind formulation stems from its success, along with the Wilson action, in classical simulations. 

Similarly, the use of the gauge-invariant Hilbert space is often regarded as essential, with the common lore saying that ``physical states must be gauge invariant.''
However, it is crucial to recognize that the gauge-invariant Hilbert space is only one of many equivalent ways to describe physical states~\cite{Hanada:2020uvt,Hanada:2021ipb,Fliss:2024don}. 
In many cases, employing non-singlet states (those that are not gauge invariant) can provide clearer insights. 
Well-established examples from classical field theory demonstrate that non-singlet descriptions can be especially useful for well-localized wave packets\footnote{
In principle, one can perform all possible gauge transformations and take a linear combination to produce a singlet state. This is a valid procedure, as nothing precludes the superposition of well-localized wave packets. Nobody does so because we do not have to use a gauge-invariant Hilbert space.
}. 
Furthermore, in connecting the Hamiltonian formulation in the gauge-invariant Hilbert space to the path integral formulation, one has to traverse the non-singlet Hilbert space (see e.g.,~Ref.\cite{Rinaldi:2021jbg}). 
The geometry emerging from matrices in string/M-theory also uses non-singlet states, often in diagonal or block-diagonal form \cite{Witten:1995im,Banks:1996vh,Hanada:2021ipb}.
Thus, while the gauge-invariant Hilbert space has its merits --- most notably, the reduction of the dimension of the Hilbert space that is crucial for simulations on \textit{classical computers} ---, adhering strictly to it may unnecessarily complicate matters, without offering clear advantages. 
Instead, it may obscure the underlying physics and hinder the efficiency of quantum simulations by dramatically increasing the complexity of the corresponding quantum circuits. 
The full potential of the gauge-invariant Hilbert space could therefore be more suited for classical computers, rather than for quantum computers.

One should note that the Kogut-Susskind Hamiltonian can be seen as a special case of the orbifold Hamiltonian 
when interpreting the unitary link variable as the infinite scalar-mass limit of the complex link variable~\cite{Bergner:2024qjl,Bergner:2025zkj}.
In practice, we can study the orbifold lattice with several values of scalar mass, and then extrapolate the results to the infinite-mass limit. 
Taking this approach allows us to leverage the universal framework. 
The term ``Kogut--Susskind Hamiltonian'' may be open to debate because different operators are used to simplify the commutation relations, but the underlying physics remains unchanged, even at the regularized level.\footnote{
Ref.~\cite{Kaplan:2002wv} introduced the orbifold lattice performing an orbifold projection to the Banks--Fishler--Shenker--Susskind (BFSS) matrix model~\cite{Banks:1996vh}, to build a supersymmetric lattice. The simplicity of the orbifold lattice Hamiltonian comes from the simplicity of the BFSS Hamiltonian. It is interesting that a technical issue associated with the complexity of the Kogut--\textit{Susskind} Hamiltonian can be resolved thanks to the simplicity of the Banks--Fishler--Shenker--\textit{Susskind} Hamiltonian.
}

Our estimate of the relationship between the lattice spacing $a$ and the necessary truncation level $\Lambda$ was qualitative. 
Even though the scaling remains the same, a more quantitative investigation may lead to the conclusion that the Kogut-Susskind Hamiltonian can be efficiently programmable on quantum computers if the value of $\Lambda$ required for sufficiently precise results turns out not to be too large.
But how can we estimate the truncation effect quantitatively? 
For this purpose, again, the orbifold lattice construction is advantageous: the truncation effect can be studied for expectation values of observables by using standard Markov Chain Monte Carlo importance sampling methods on a classical device~\cite{Hanada:2022pps}. 
By taking a large bare scalar mass, we can use the orbifold lattice Hamiltonian to study the truncation effect in the Kogut-Susskind Hamiltonian, too.

Given the rapid development of quantum computers, it is crucial to establish a theoretically sound framework for quantum simulations that scales efficiently.
While we advocate the power of a simple, universal, and concrete framework that abstracts away model-specific details, it would be interesting to explore whether theory-specific approaches can offer competitive advantages. 
In the case of systems of bosons, any such advantages would need to, at a minimum, match or exceed the reduction in circuit length discussed in this paper.

\begin{center}
\section*{Acknowledgment}
\end{center}
The authors thank Georg Bergner, Anthony Ciavarella, Olivia Di Matteo, Dorota Grabowska, Saurabh Kadam, Angus Kan, Michael Kreshchuk, Lin Lin, Mason Rhodes, Claudio Sanavio, David Schaich, Artur Scherer, Jesse Stryker, Simone Romiti, Carsten Urbach, Graham Van Goffrier, Yoshimasa Hidaka, and Xiaoyang Wang for useful discussions.
The authors also thank Maria Tudorovskaya, Frederic Sauvage, Yuta Kikuchi, and Andreas Sch\"{a}fer for comments on the manuscript.
M.~H. and E.~R. thank the Royal Society International Exchanges award IEC/R3/213026.
M.~H.~thanks the STFC for the support through the consolidated grant ST/Z001072/1.
E.~M.~thanks UK Research and Innovation Future Leader Fellowship MR/S015418/1, MR/X015157/1, and the STFC grants ST/T000988/1, ST/X000699/1.

\appendix
\section{Cost comparison of lattice Hamiltonian formulations of Yang-Mills theories}\label{appendix:cost_analyses}
In this appendix, we analyze the cost scaling for quantum simulation of Yang-Mills theory using several formulations not addressed in the main text.\footnote{
We thank Anthony Ciavarella, Dorota Grabowska, Yoshimasa Hidaka, Saurabh Kadam, and Jesse Stryker for clarification of their work.} 
In contrast to previous studies, which typically focus on overall qubit count or lattice volume scaling, we examine how circuit complexity scales with the number of qubits $Q$ per bosonic degree of freedom. 
Specifically, we investigate the worst-case scaling behavior for both circuit depth and compilation cost as functions of $Q$. 
In all cases we investigate, there is no systematic analysis because the Hamiltonian is complicated. 
At present, there is little evidence suggesting that exponential scaling in simulation cost can be avoided in the following approaches.

\subsection{Gauge fixing approach}
One approach to reducing simulation costs is to employ a gauge-fixed version of the Kogut-Susskind Hamiltonian, as demonstrated in recent studies of SU(2) theory in the maximal tree gauge; see refs.~\cite{DAndrea:2023qnr,Grabowska:2024emw}. 
Gauge fixing proves particularly effective for one-dimensional spatial systems, where the gauge field—and consequently the link variable—can be completely eliminated. 
However, this advantage disappears in higher dimensions. 
When multiple spatial dimensions are present, the number of link variables that cannot be eliminated scales proportionally with the spatial volume, reproducing the same computational complexity as the original formulation without gauge fixing.

An additional complication arises from the nonlocality introduced by gauge fixing, which may significantly increase the complexity of the circuit design. 
This mirrors the challenges encountered with gauge-invariant Hilbert spaces discussed in Sec.~\ref{sec:gauge_invariance}. 
At present, no known methods exist to circumvent the exponentially scaling resource requirements, whether measured in terms of circuit depth or compilation overhead.
Moreover, there is no available circuit implementation of the time evolution of a $(3+1)$-dimensional non-Abelian lattice gauge theory, which is instead easily provided in the orbifold formalism.

\subsection{Loop-String-Hadron formulation}
The Loop-String-Hadron (LSH) formulation~\cite{Raychowdhury:2019iki} aims to reduce the gauge redundancy. 
Although this has certain advantages in describing gauge-invariant physics, the Hamiltonian necessarily becomes complicated, and as a consequence, it is difficult to design quantum circuits. 
At present, there is no systematic resource estimate beyond the SU(2) theory in one spatial dimension~\cite{Davoudi:2022xmb}.
Some progress on defining the Hamiltonian for SU(3) in the LSH formalism has been made in one and two spatial dimensions~\cite{Kadam:2022ipf, Kadam:2024ifg}, but no circuit representation has been derived due to the complicated Hamiltonian.
However, the general formalism of Ref.~\cite{Davoudi:2022xmb} should be implementable for any non-Abelian gauge group and in any spatial dimension, leading to an, in principle, exponential improvement in number of gates with respect to naive approaches.
Quantum circuits with an exponential improvement in this formulation will still require to work out some ingredients of the formalism that are very labor intensive (e.g. all the matrix elements of the plaquette operator in the truncated basis of choice) and are accompanied by a large pre-factor when it comes to classical computational time, albeit not exponential).

\subsection{$1/N$ expansion}
Recent work~\cite{Ciavarella:2024fzw,Ciavarella:2025bsg} proposed utilizing the $1/N$ expansion to simplify quantum simulation of the Kogut-Susskind Hamiltonian.
This approach exploits a well-known property of the 't~Hooft large-$N$ limit: in the confined phase, properly normalized short strings (Wilson loops) form an orthonormal basis up to $1/N$ corrections, while string interactions (joining and splitting induced by the electric term of the Hamiltonian) are suppressed by $1/N$. 
However, this approach faces fundamental limitations in several important and relevant scenarios:
\begin{itemize}
    \item Typical high-energy states, particularly those in the deconfined phase, involve long strings with numerous (self-)intersections whose length scales with $N$. These states do not form an orthonormal basis, and interactions remain significant. Although the contribution from each intersection is $1/N$-suppressed, the total contribution is not suppressed due to the proliferation of intersections (see, e.g., Ref.~\cite{Hanada:2014noa}).
    
    \item The $1/N$ expansion is asymptotic in nature, making the systematic study of certain nonperturbative phenomena challenging.
    
    \item Incorporating $1/N$-suppressed terms requires intricate combinatorics in loop space, significantly complicating extensions to groups like SU(3).
\end{itemize}

The complexity becomes apparent already at next-to-leading order in the $1/N$ expansion, as observed in Ref.~\cite{Ciavarella:2025bsg}. 
Consequently, no systematic analysis of simulation costs exists for arbitrary truncation levels and for SU(3).
Some numerical evidence exists that, in some region of parameter space, only a small truncation value and a low order in the $1/N$ expansion are needed to reproduce SU(3) physical properties such as the glueball mass in two spatial dimensions.
Without more systematic studies, there is no reason to expect that exponential scaling in simulation cost can be avoided.

\subsection{$q$-deformation}
Refs.~\cite{Zache:2023dko,Hayata:2023bgh} used $q$-deformed Lie algebra to regularize the infinite-dimensional Hilbert space of the Kogut-Susskind formulation. 
Despite mathematical elegance, when it comes to practical issues concerning quantum simulations, the Hamiltonian is too complicated.
Hence, there is no estimate for how the digitization errors will scale for SU(3) in three spatial dimensions. 
Currently, nothing indicates that the exponential scaling in simulation cost can be avoided.

\bibliographystyle{utphys}
\bibliography{reference}

\end{document}